\documentstyle[aps,pra,subfigure,epsfig,amsmath,amssymb]{revtex}

\begin{document}
\draft

\title{Momentum flux density, kinetic energy density and their
  fluctuations for one-dimensional  
  confined gases of non-interacting fermions}

\author{Anna Minguzzi, Patrizia Vignolo and M.~P. Tosi}
\address{Istituto Nazionale di Fisica della Materia and Classe di Scienze,
Scuola Normale Superiore,
Piazza dei Cavalieri 7, I-56126 Pisa, Italy}
\maketitle
\begin{abstract}
We present a Green's function method for the evaluation of the particle 
density profile and of the higher moments of the one-body density matrix in 
a mesoscopic system of $N$ Fermi particles moving independently in a
linear potential. 
The usefulness of the method is illustrated by applications to a Fermi gas
confined in a harmonic potential well, for which we evaluate
the  momentum flux  and  kinetic energy densities  as well as  their 
quantal mean-square fluctuations. 
We also study some properties of 
the kinetic energy functional $E_{kin}[n(x)]$ in the same system. 
Whereas a local approximation to the kinetic energy  
density yields a multi-valued function,
 an exact single-valued relationship between the density 
derivative of $E_{kin}[n(x)]$ and the particle density $n(x)$ is demonstrated
and evaluated for various 
values of the number of particles in the system. 
\end{abstract}
\pacs{PACS numbers: 03.75.Fi, 05.30.Fk, 31.15.Ew}

\section{Introduction} 
The study of ultracold gases of fermionic alkali atoms is being pursued
experimentally \cite{fermioni} by using basically the same trapping and 
cooling techniques which have led in 1995 to the achievement of Bose-Einstein 
condensation in gases of bosonic isotopes \cite{bec_rb,bec_na,bec_li}.
The main focus of the experimental effort is towards the realization of novel
superfluids from pairing between fermions in different hyperfine states.
However, attention has also been given to dilute Fermi gases in a fully 
spin-polarized state under magnetic confinement. Such a system presents 
special motives of fundamental interest from the fact that the Pauli 
principle suppresses the interactions in the {\it s}-wave scattering
channel and thereby essentially isolates the kinetic energy functional
which is invoked in the density functional theory of inhomogeneous fluids
 \cite{lund}.  

A quasi-one dimensional Fermi gas can be realized in the experiments
 \cite{fermioni} inside a very anisotropic magnetic trap, producing
approximately harmonic confinement which is very tight in two
directions.  The particle and kinetic energy densities of the
corresponding one-dimensional (1D) model of non-interacting fermions
in a harmonic potential well have been evaluated exactly up to
mesoscopic numbers of particles \cite{noi}. The shell structure in the
particle density which was noticed in earlier studies of 3D Fermi
gases \cite{schneider,bruun} is greatly enhanced in lower
dimensionality (see also Ref.~\cite{zimmermann}).
As a further motive of interest, the 1D
model of noninteracting fermions has the same particle density profile
and thermodynamic properties as the 1D model of point-like bosons
with hard core interactions \cite{girardeau,girardnew,kolomeisky}.
The system of interacting bosons in 1D is attracting attention due to
its rich phase diagram \cite{shlyap}.

The particle and kinetic energy densities are the zero-th moment and
 a second moment of the one-body density matrix. In our earlier
study~ \cite{noi} we have outlined a general method by which one may
calculate 
the successive moments of the density matrix in a system of $N$ Fermi
particles moving independently in a 1D potential. 
 In the fully quantum regime alternative definitions can be given for
the  higher-order moments, depending on how the
operators are symmetrized~ \cite{uhlenbeck}. In particular, two
 different second-order moments can be defined for an inhomogeneous
 gas under confinement, yielding two different physical quantities 
which are the kinetic energy density and the momentum flux density.
 The method  here proposed is suitable for the higher-order
moments  in any of their forms. It
exploits
the definition of the Green's function operator $\hat G(x)$ in coordinate
space and allows one to use the numerical methods already developed in
solid state physics \cite{vfg,fvg,fgv} to evaluate the density
of single-particle 
states from the Green's function in the energy domain.

In this paper we give a detailed account of the general method in
Sec.~\ref{sec1} and some new applications to the case of harmonic
confinement in the following Sections.  In Sec.~\ref{secT_Pi} we
contrast the kinetic energy density with the momentum flux density
in the presence of inhomogeneity.
 In Sec.~\ref{secflutt} we assess the quantal
(zero-temperature) fluctuations in the  kinetic energy and 
momentum flux densities by means of
a calculation of  fourth moments of the one-body density
matrix. Properties of the kinetic energy functional are discussed in
Sec.~\ref{secfunc} in the context of density functional theory. 
An exact local-density method exists for the system of present interest, as
shown in a different approach by Lawes and March \cite{march}, and is
explicitely exhibited here for various values of the number of Fermi
particles. Finally, in Sec.~\ref{secconcl} we summarize our main
conclusions.

\section{The method}
\label{sec1}
We consider a system of $N$ non-interacting Fermi particles in one spatial
dimension subjected to an external confinement given by a real
potential $V_{ext}(x)$. We may take the
eigenfunctions $\psi_i(x)$ as real~ \cite{noticina}.
The 
one-body density matrix at zero temperature is
\begin{equation}
\rho(x,x_1)=\sum_{i=1}^N\psi_i(x)\psi_i(x_1).
\label{rho_def}
\end{equation}
Its zero-th moment is the
equilibrium density profile $n(x)$, and its odd moments are zero due to the
 reality of the wavefunctions.
For the even moments of higher order several definitions have been
considered  in
the literature 
(for a review see {\it e.g.} Ziff {\it et al.}~ \cite{uhlenbeck}), among
which we shall 
focus on the following:
\begin{equation}
P_n(x)=(-i)^n\left[\dfrac{\partial^n}{\partial
x_1^n}\rho(x,x_1)\right]_{x_1=x}
\label{stella}
\end{equation}
and
\begin{equation}
S_n(x)=(-i)^n\left[\dfrac{\partial^n}{\partial
r^n}\rho(x+r/2,x-r/2)\right]_{r=0}
\label{stellastella}
\end{equation}
 (we have set $\hbar=1$ and $m=1$).
For $n\ge 2$ these definitions give rise  to
different equilibrium properties in an inhomogeneous system, although
their semiclassical limits and their integrals 
coincide. 
In the specific case of $n=2$ Eqs. (\ref{stella}) and
(\ref{stellastella})
give  twice the  kinetic energy density and the momentum flux
density, respectively:  these will be discussed in Sec.~\ref{secT_Pi}.

The method developed in Ref.~\cite{noi} is here extended to evaluate both
$P_n(x)$ and $S_n(x)$. For this purpose we  write
Eqs.~(\ref{stella}) and (\ref{stellastella}) in terms of the Green's
function $\hat G(x)=(x-\hat{x}+ i \varepsilon)^{-1}$ in coordinate space as
\begin{equation}
P_n(x)=-\frac{1}{\pi}\lim_{\varepsilon\rightarrow0^+} {\rm
Im}\,\sum_{i=1}^N \langle \psi_i\,|\hat G(x) \hat p^n |\,\psi_i\rangle\;
\label{equno}
\end{equation}
and
\begin{equation}
S_n(x)=-\frac{1}{2^n\pi}\sum_{\sigma=0}^n\left(\begin{array}{c}
n\\\sigma\\\end{array}\right)\lim_{\varepsilon\rightarrow0^+}
{\rm Im}\,\sum_{i=1}^N \langle \psi_i\,|\hat{p}^{\sigma}\hat G(x)\hat
p^{n-\sigma} |\,\psi_i\rangle\;.
\label{eqdue}
\end{equation}
Here, $|\,\psi_i\rangle$ are the eigenstates of the Hamiltonian in the
coordinate representation and $\hat p$ is the momentum operator. 
Equation~(\ref{equno})
is easily proved by making use of the expression of $\hat G(x)$
in the coordinate representation:
\begin{eqnarray}
P_n(x)&=&-\frac{(-i)^n}{\pi}\lim_{\varepsilon\rightarrow0^+} {\rm
Im\,}\sum_{i=1}^N \int {\rm d}x_i
\psi_i(x_i)\frac{1}{x-x_i+i \varepsilon}\; \dfrac{\partial^n}{\partial
x_i^n}\psi_i(x_i)\nonumber \\
&=&{(-i)^n}\sum_{i=1}^N \int {\rm d}x_i
\psi_i(x_i)\delta(x-x_i)\dfrac{\partial^n}{\partial x_i^n}\psi_i(x_i)\;,
\end{eqnarray}
where $x_i$ are the eigenvalues of the position operator $\hat{x}$.
Equation (\ref{eqdue}) can be proved in a similar fashion.

The momentum distribution $n(p)$ can be  analogously written  in terms of the trace of  the Green's function
$\hat G(p)=(p-\hat p + i\varepsilon )^{-1}$ in  momentum space, taken  on the
eigenvectors $|\,\phi_i\rangle$ of the Hamiltonian in the 
momentum representation:
\begin{equation}
n(p)=
\sum_{i=1}^N \int {\rm d}p_i
|\phi_i(p_i)|^2\delta(p-p_i)
\end{equation}
or
\begin{equation}
n(p)=-\frac{1}{\pi}\lim_{\varepsilon\rightarrow0^+} {\rm
Im}\,\sum_{i=1}^N \langle \phi_i\,|\hat G(p) |\,\phi_i\rangle\;,
\end{equation}
where $p_i$ are the eigenvalues of the momentum operator.

To evaluate the Green's functions in the specific case of harmonic
confinement,  
we make use of the representation for
the position and the momentum operators 
in the basis  of the eigenstates of the harmonic oscillator, {\it i.e.}
$\hat{x}=(a+a^{\dag})/\sqrt 2$ and
$\hat{p} =i(a^{\dag}-a)/\sqrt 2$
with $a\,|\,\psi_i\rangle=\sqrt{i-1}\,|\,\psi_{i-1}\rangle$ and
$a^{\dag}\,|\,\psi_i\rangle=\sqrt{i}\,|\,\psi_{i+1}\rangle$ (we have
set the harmonic oscillator frequency $\omega=1$).
Explicitly, the representation of $\hat x$ in matrix form 
is given by
\begin{equation}
{\hat{x}}=\left(\begin{array}{ccccccc}
         0& 1/\sqrt{2}&     & & & &\\
        1/\sqrt{2}& 0& 1& &  & &\\
             & 1& 0& \sqrt{3/2}&  & &\\
             & &\sqrt{3/2}& 0&\sqrt{2}& & \\
             &   &         & \sqrt{2}  & 0& \sqrt{5/2} &\\
             &   &         &  & \ddots& \ddots & \ddots\\
\end{array}\right)
\end{equation}
and similarly for $\hat p$.
The various  moments of the one-body density matrix and the momentum
distribution are then evaluated as 
the trace of the operators $\hat
p^\sigma\hat G(x) \hat p^{n-\sigma}$ and of $\hat G(p)$  on the first $N$
states (Tr$_N$).
The trace is calculated by the general method given below. The
quantities $n(x)$ and $n(p)$ could also be calculated from 
 a suitably 
modified Kirkman-Pendry relation \cite{pendry,fgv}, as is  shown in 
 Appendix A.

\subsection{Expression for the evaluation of the trace}

The partial trace ${\rm Tr}_N$ of a generic matrix $Q$ is related to
the determinant of the inverse matrix $Q^{-1}$ by
\begin{equation}
{\rm Tr}_NQ=\partial
\left[\ln\det(Q^{-1}+\lambda{\Bbb{I}}_N)\right]/\partial\lambda|_{\lambda=0}.
\label{chefatica}
\end{equation}
This relation  is demonstrated as follows:
\begin{eqnarray}
\partial
\left[\ln\det(Q^{-1}+\lambda{\Bbb{I}}_N)\right]/\partial\lambda|_{\lambda=0}=
\partial
\left[ {\rm Tr} \ln (Q^{-1}+\lambda{\Bbb{I}}_N
)\right]/\partial\lambda|_{\lambda=0}\nonumber \\
={\rm Tr} \left[(Q^{-1}+\lambda{\Bbb{I}}_N)^{-1}{\Bbb{I}}_N
\right]_{\lambda=0}={\rm Tr}(Q{\Bbb{I}}_N)={\rm Tr}_N Q.
\end{eqnarray}

Here we take $Q=\hat p^\sigma \hat G(x) \hat p^{n-\sigma}$. In
the case of harmonic confinement 
this matrix     
is the product of a $(2n+1)$-diagonal matrix and of the
inverse of a tridiagonal matrix. 
Exploiting Eq.~(\ref{chefatica}) we can write
\begin{equation}
P_n(x)=-\frac{1}{\pi}\lim_{\varepsilon\rightarrow0^+}{\rm Im}
\frac{\partial}{\partial\lambda}
\left[\ln\det(x+i\varepsilon-\hat{K}^{n,0})
\right]_{\lambda=0},
\label{penta}
\end{equation}
and
\begin{equation}
S_n(x)=-\frac{1}{2^n\pi}\sum_{\sigma=0}^n\left(\begin{array}{c}n
\\\sigma\\\end{array}\right)\lim_{\varepsilon\rightarrow0^+}{\rm Im}
\frac{\partial}{\partial\lambda}
\left[\ln\det(x+i\varepsilon-\hat{K}^{n,\sigma})
\right]_{\lambda=0}
\end{equation}
where $\hat K^{n,\sigma}=\hat{x}-\hat
p^{n-\sigma}\lambda{\Bbb{I}}_N\hat{p}^{\sigma}$.
We have thus reduced the problem to the evaluation of the determinant
of the matrix $(x+i\varepsilon-\hat{K}^{n,\sigma})$,
which is  tridiagonal in the tail and whose first $N$ rows
have only a few non-vanishing elements for low values of $n$.

If we make a partition in blocks of dimension $n\times n$  for the
first part of the matrix,
the whole operator can be written as
\begin{equation}
\hat{K}^{n,\sigma}=
\lim_{M\rightarrow\infty}\hat{K}^{n,\sigma}_M=
\lim_{M\rightarrow\infty}\left(\begin{array}{c c c c c l}
 {\cal   A}_{1} & {\cal B}_{1,2} &     & & &\\
 {\cal  B}_{2,1} & {\cal A}_{2}&{\cal  B}_{2,3} & &  &\\
 &{\cal B}_{3,2} & \ddots &{\cal  B}_{3,4} &  &\\
         &  & \ddots & \ddots & \ddots &\\
	&&  & \ddots & \ddots &{\cal B}_{M\!-\!1,M}\\ 
       &  &  &   &{\cal  B}_{M,M\!-\!1}&{\cal A}_M\\
\end{array}\right).\label{tridiag}
\end{equation}
Here, we have introduced
 the truncated matrix $\hat{K}^{n,\sigma}_M$, which will be
used below in actual numerical calculations.

\subsection{The renormalization procedure}
The determinant of the matrix $(x+i\varepsilon-\hat{K}^{n,\sigma})$ 
with a tridiagonal representation
such as that in Eq.~(\ref{tridiag}) can be factorized into
a product of determinants
of matrices having the same dimension as that  of the blocks
of the partition \cite{fgv}.

As a first step, it is easy to show that if we 
partition the matrix $\hat{K}^{n,\sigma}_M$ into four blocks,
thereby defining $\tilde{\cal B}_{i,j}$ an  $\hat{K}^{n,\sigma}_i$ from
\begin{equation}
\label{dieci}
\hat{K}^{n,\sigma}_M=\left(\begin{array}{cc}
        \hat{K}^{n,\sigma}_{M\!-\!1} & \tilde{\cal B}_{M\!-\!1,M} \\
        \tilde{\cal B}_{M,M\!-\!1} &{\cal A}_M \\
\end{array}\right)\;,
\end{equation} 
we can write
\begin{equation}
\det(x+i\varepsilon-\hat{K}^{n,\sigma}_M)=\det(x+i\varepsilon-\hat{K}^{n,\sigma}_{M-1})
\cdot\det(x+i\varepsilon-{\cal A}_M-
\tilde{\cal B}_{M,M\!-\!1}
(x+i\varepsilon-\hat{K}^{n,\sigma}_{M\!-\!1})^{-1}\tilde{\cal B}_{M\!-\!1,M}).
\end{equation}
Applying recursively  this procedure and taking the limit
$M\rightarrow\infty$ we obtain
\begin{equation}
\det{(x+i\varepsilon-\hat{K}^{n,\sigma})}=
\det {(x+i\varepsilon-{\cal A}_{1})}\cdot \lim_{M\rightarrow\infty}\prod_{j=2}^{M}
\det {(x+i\varepsilon-{\cal A}_j \!-\!\tilde{\cal B}_{j,j\!-\!1}
(x+i\varepsilon-\hat K^{n,\sigma}_{j\!-\!1})^{\!-\!1}\tilde{\cal
B}_{j\!-\!1,j})}\,.
\label{PAPPA}
\end{equation}

It is now important to notice that, owing to the particular form of
the matrices  
$\tilde{\cal B}_{j\!-\!1,j}$
and $\tilde{\cal B}_{j,j\!-\!1}$,
it is not necessary to explicitly invert the matrices 
$(x+i\varepsilon-\hat K^{n,\sigma}_{j\!-\!1})$.
Rather, 
we may  use a renormalization procedure for the operator $\hat
K^{n,\sigma}_{j\!-\!1}$ 
to further simplify the expression (\ref{PAPPA}) and to 
calculate the inverse of matrices having dimension
at most equal to 
$n\times n$. 
Renormalization is based on the idea  that
if one only needs to describe a given subspace of the whole Hilbert
space of the system, one can define an effective operator acting in
the subspace and taking  into account the contribution from the rest of
the system. Renormalization allows us to write
\begin{equation}
\det(x+i\varepsilon-\hat{K}^{n,\sigma})=
\prod_{j=1}^{\infty}\det(x+i\varepsilon-\tilde{\cal A}_j)
\label{INFM}
\end{equation}
where $\tilde{\cal A}_1={\cal A}_1$ and
\begin{equation}
\tilde{\cal A}_j={\cal A}_j+{\cal B}_{j,j-1}
(x+i\varepsilon-\tilde{\cal A}_{j-1})^{-1}{\cal B}_{j-1,j}
\label{ric}
\end{equation}
for $j>1$. The ${\cal B}_{j, j+1}$
have been introduced in Eq.~(\ref{tridiag}).
In the specific case of $n=2$ and  $\sigma=0$ Eq.~(\ref{INFM}) yields
back the expression obtained in Ref.~\cite{noi}.

Operatively,  we have studied the convergence of the function
$\det(x+i\varepsilon-\hat{K}^{n,\sigma}_M)$ on increasing $M$. We
have performed our  
calculations up to $n=4$ and $N=1000$  fermions with $M\sim10^7$ and
$\varepsilon=0.001$. 

\section{Kinetic energy and momentum flux densities}
\label{secT_Pi}

The second moments $P_2(x)$ and $S_2(x)$ of the one-body density
matrix from Eqs.~(\ref{stella})
and~(\ref{stellastella}) have different physical meanings and
 show different behaviours. The function $P_2(x)=\sum_i^N \langle
\psi_i| \delta(x-x_i) \hat p^2 | \psi_i \rangle$ is  twice
the  kinetic energy density $T(x)$  \cite{march} and for
harmonic confinement  has already been
evaluated in Ref.~\cite{noi} together with the equilibrium density
profile. This quantity is of main interest in the context of Density Functional
Theory, as will be discussed in Sec.~\ref{secfunc}. On the other hand,
$S_2(x)=(1/2) \sum_i^N \langle
\psi_i| \delta(x-x_i) \hat p^2 +\hat p\, \delta(x-x_i)\, \hat p| \psi_i
\rangle$ is the momentum flux density $\Pi(x)$ which enters
the equations of generalized hydrodynamics
 \cite{march_tosi}. In 3D fluids this quantity becomes a tensor, known
as the
kinetic stress tensor. 

The equation of motion of the
particle density $n(x,t)$ for
non-interacting fermions in 1D is
\begin{equation}
\frac{\partial^2}{\partial t^2}n(x,t)= \frac{\partial^2}{\partial x^2}
\Pi(x,t) + \frac{\partial}{\partial
x}\left[n(x,t)\frac{\partial}{\partial x}V_{ext}(x,t) \right].
\label{bessie}
\end{equation}
At equilibrium in a static potential 
we obtain from Eq.~(\ref{bessie}) an exact relation
between the momentum flux density and the density profile,
\begin{equation}
\frac{d}{d x} \Pi(x)= -n(x) \frac{d}{d
x}V_{ext}(x) +{\rm const} \;.
\end{equation}
The integration constant can be fixed by imposing suitable
boundary conditions for $x\rightarrow \pm \infty$.

Another exact relation can be derived by exploiting
Eqs.~(\ref{rho_def})-(\ref{stellastella}) together with the definition
of $n(x)$ as the zero-th moment of the density matrix. This 
reads:
\begin{equation}
\Pi(x) = 2 T(x) +\frac{1}{4} \frac{d^2}{d x^2} n(x)\;.
\label{patty}
\end{equation}
Equation~(\ref{patty}) exhibits the relation which exists between the momentum
flux density and the kinetic pressure $P(x)=2 T(x)$ in the inhomogeneuos
system. Evidently, in the
homogeneous limit the momentum flux density and the kinetic pressure
coincide -- and therefore they also coincide
within a local-density description for the confined gas.
They instead differ within a
quantum-mechanical description of the
the inhomogeneous system, when one keeps exactly into account  the
role of confinement. 
This difference is readily illustrated in the fully quantum case of a
single  fermion
in an harmonic oscillator potential, where we have 
\begin{equation}
\Pi(x)=T(x)+V(x)
\end{equation}
with
$V(x)=n(x) V_{ext}(x)$
being the potential energy density associated with the confinement.

\begin{figure}
\mbox{
\subfigure[]
{\epsfig{figure=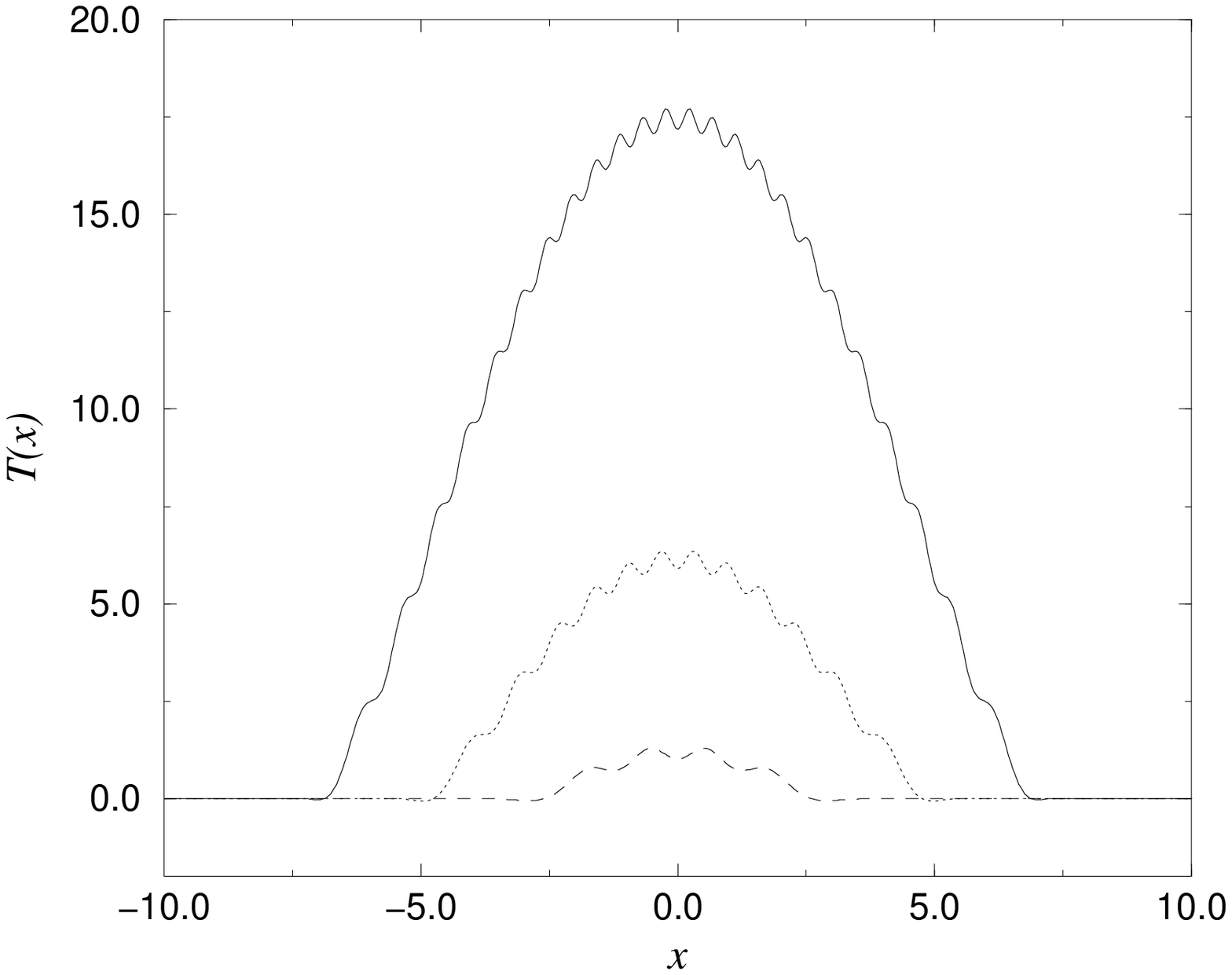,width=0.47\linewidth}}
\subfigure[]
{\epsfig{figure=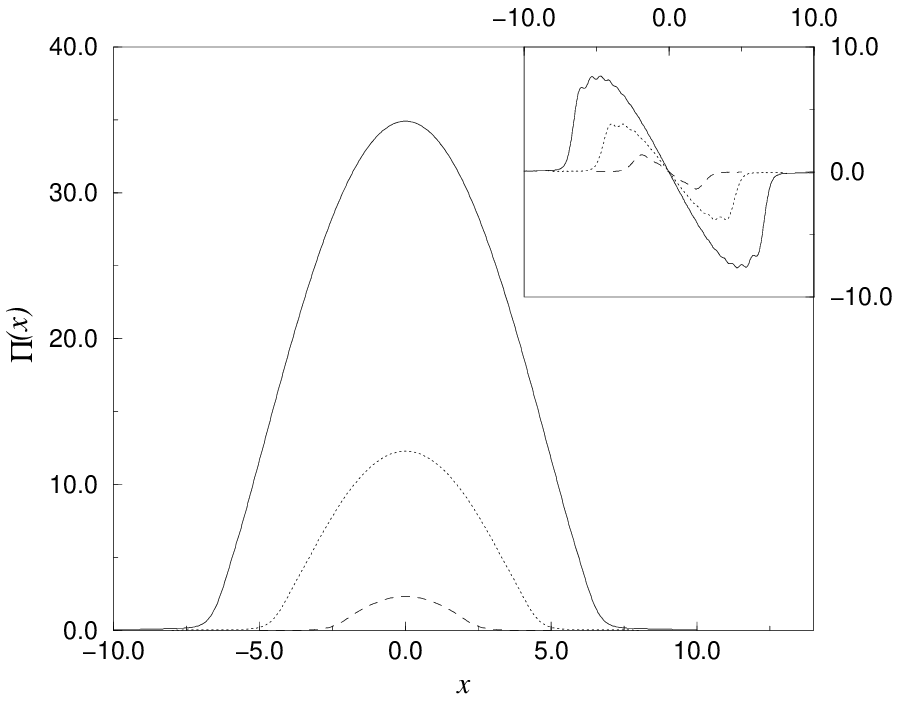,width=0.47\linewidth,height=0.40\linewidth}}}
\caption{Exact kinetic energy density $T(x)$ (a) and  momentum flux
density $\Pi(x)$
(b),  in units of  $\hbar
\omega/a_{ho}$
 as 
functions of the spatial coordinate $x$ (in units of $a_{ho}=\sqrt{\hbar/m\omega}$)
for $N$ = 4 fermions (dashed line), 12 fermions (dotted line) and 24
fermions  (solid line) 
in a 1D
harmonic potential. The
inset in (b) shows the  derivative $d \Pi(x)/dx$.}
\label{figT_Pi}
\end{figure}

For the case of  harmonic confinement we have employed the method
described in Sec.~\ref{sec1} to evaluate the exact profiles for both $T(x)$
and $\Pi(x)$ at various numbers of particles. Figure~\ref{figT_Pi}
shows these results for $N$= 4, 12 and 24 fermions. While $T(x)$ shows $N$
oscillations and negative tails~ \cite{noi}, $\Pi(x)$ is everywhere
positive and its shell structure is visible only in its first derivative
(see inset in Fig.~\ref{figT_Pi}).

\section{Quantum
fluctuations of the kinetic energy  and momentum flux densities}
\label{secflutt}

From the fourth moments 
$S_4(x)$ and $P_4(x)$ of the density matrix
we can evaluate the local fluctuations of the
momentum flux  and kinetic energy densities.
For the former  the fluctuations are most conveniently
introduced through the Wigner 
distribution function,
\begin{equation}
f_W(p,x)=\int dr\, e^{ipr} \rho(x+r/2,x-r/2)\;.
\end{equation}
In this representation the relevant  moments are given by $n(x)=\sum_p
f_W(p,x)$ 
and by $\Pi(x)=\sum_p p^2 f_W(p,x)\equiv n(x) \langle p^2 \rangle$. The
mean square fluctuation of the momentum flux density is therefore
given by 
\begin{equation}
\Delta \Pi(x)=\sum_p (p^2-\langle p^2\rangle)^2
f_W(p,x)=S_4(x)-{\Pi^2(x)}/n(x) 
\;,
\label{quadr}
\end{equation}
where $S_4(x)$ is as defined in
Eq.~(\ref{stellastella}). Equation~(\ref{quadr}) in the coordinate
representation for the density matrix can be rewritten as
\begin{equation}
\Delta \Pi(x)=\left[-\dfrac{\partial^2}{\partial
r^2}-\dfrac{\Pi(x)}{n(x)} \right]^2  \left.\rho(x+r/2,x-r/2)\right|_{r=0}
\;.
\end{equation}
The fluctuations of the kinetic energy density  are 
obtained in a similar 
fashion as 
\begin{eqnarray}
\Delta T(x)&=&\left[-\dfrac{1}{2}\dfrac{\partial^2}{\partial
x_1^2}-\dfrac{T(x_1)}{n(x_1)} \right]^2 \left.
\rho(x,x_1)\right|_{x_1=x}\nonumber 
\\ &=& \dfrac{1}{4} P_4(x) - \dfrac{T^2(x)}{n(x)} +\dfrac{1}{2}
\dfrac{d}{dx}\left[ n(x)\dfrac{d}{dx}\left(\dfrac{T(x)}{n(x)} \right)\right]
\;,
\end{eqnarray}
with $P_4(x)$ as defined in Eq.~(\ref{stella}). 

For harmonic confinement in the local density approximation (LDA)  we have
$n_{LDA}(x)=(2N-x^2)^{1/2}/\pi$, $\Pi_{LDA}(x)=2\,
T_{LDA}(x)=(2N-x^2)^{3/2}/3\pi$ and
$S_4^{LDA}(x)=P_4^{LDA}(x)=(2N-x^2)^{5/2}/5\pi$. It is straightforward
to show from these expressions that the integrated mean square fluctuations,
divided by the square of the  corresponding integrated quantity, scale
as $1/N$. 
The same scaling is found in the exact numerical results at
large $N$.

\begin{figure}
\mbox{
\subfigure[]
{\epsfig{figure=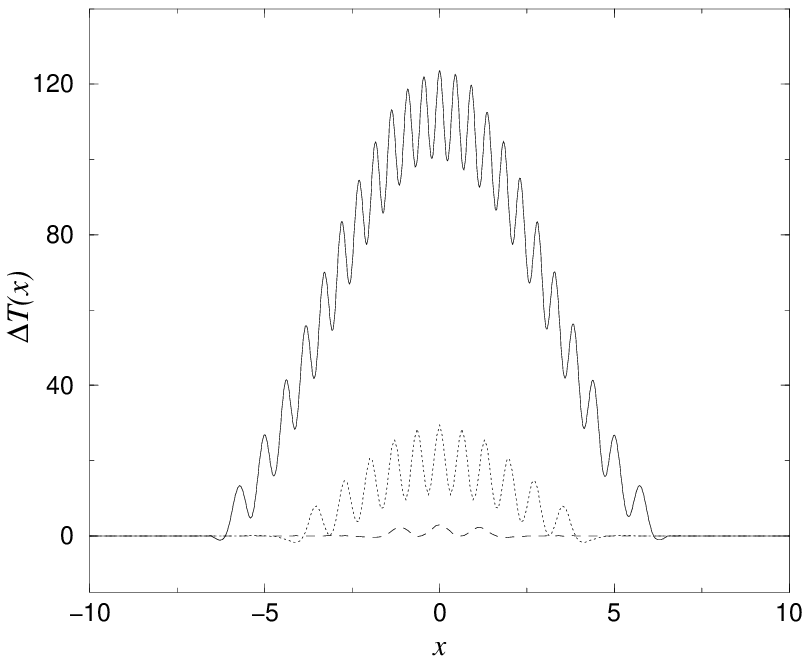,width=0.47\linewidth}}
\subfigure[]
{\epsfig{figure=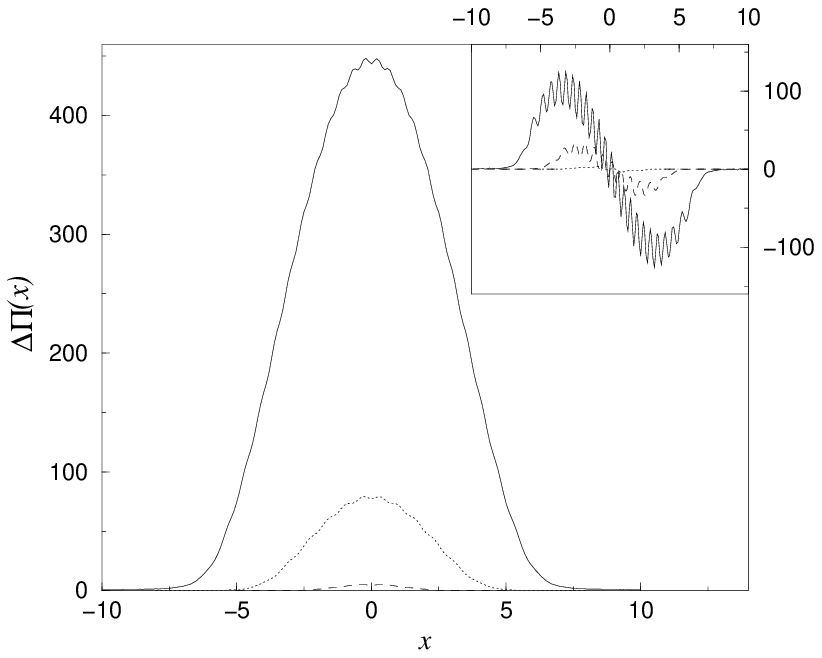,width=0.47\linewidth,height=0.42\linewidth}}}
\caption{Quantal fluctuations of the  kinetic energy density (a) and
of the  momentum flux density (b)
(in units of $(\hbar \omega)^2/a_{ho}$), as  functions of the spatial
coordinate $x$ (in units of $a_{ho}$). The numbers of particles and
the symbols are the same as in Fig.~\ref{figT_Pi}.
The
inset in (b) shows the first derivative $d \Delta \Pi(x)/dx$.}
\label{fig_flutt}
\end{figure}

\begin{figure}
\epsfig{file=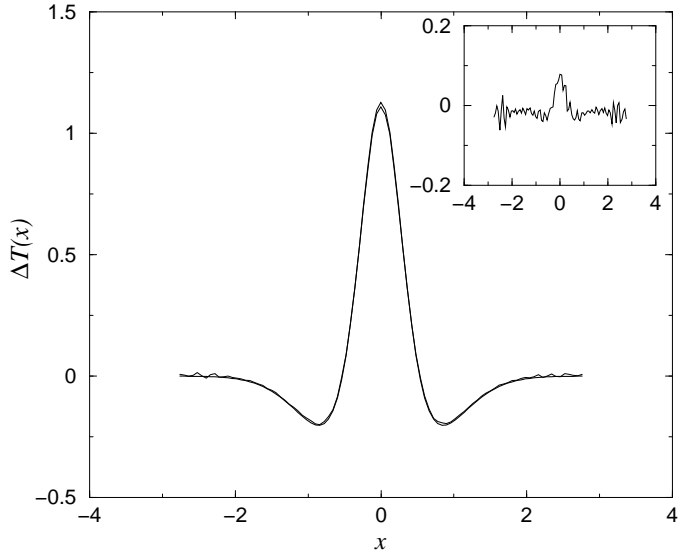,width=0.5\linewidth}
\caption{Quantal fluctuations  $\Delta T(x)$ of the kinetic
energy density   for a system of two fermions
evaluated numerically by the Green's 
function method (dotted line), compared with the result of an analytic
calculation 
(continuous line). The difference between the two curves is shown in
the inset. The units are the same as in Fig.~\ref{fig_flutt}.}
\label{ennedue}
\end{figure}

We have used the Green's function method described in
Sec.~\ref{sec1} to evaluate  the
exact profiles for both $\Delta \Pi(x)$ and $\Delta T(x)$
in harmonic confinement, 
as shown in Fig.~\ref{fig_flutt} for $N$ =4, 12 and 24 fermions.
The
positions of the $N-1$ maxima in $\Delta T(x)$  are located in
correspondence to the minima of the kinetic energy density in
Fig.~\ref{figT_Pi} (a), and again negative tails are present at the
boundaries.  
In the case of $N=2$ we have tested our numerical results against the
analytic expression given for $\Delta T(x)$ by low-order  Hermite
polynomials, the result being  shown in
Fig.~\ref{ennedue}. Evidently, we have obtained in this case a
very accurate numerical evaluation of up to the fourth moment of
the density matrix.

\section{Properties of the kinetic energy functional}
\label{secfunc}

We have seen that the  second moment $P_2(x)$ of the one-body density
matrix yields the  kinetic energy density
$T(x)=P_2(x)/2$. The {\it total} kinetic 
energy is  obtained by integration over the spatial coordinate and in
the spirit of 
Density-Functional Theory (DFT)  \cite{lund} can
be viewed as a functional of the particle density $n(x)$:
\begin{equation}
E_{kin}[n(x)]=\int dx\,T(x)=\int dx\,\hat T[n(x)]\;.
\end{equation}
Here we have introduced an operator $\hat T$ which acts on
the density profile to yield 
\begin{equation}
\hat T[n(x)]=T(x)\;.
\end{equation}
A main point of interest 
 for DFT is to assess the properties of the unknown operator
 $\hat T$.

\begin{figure}
\epsfig{file=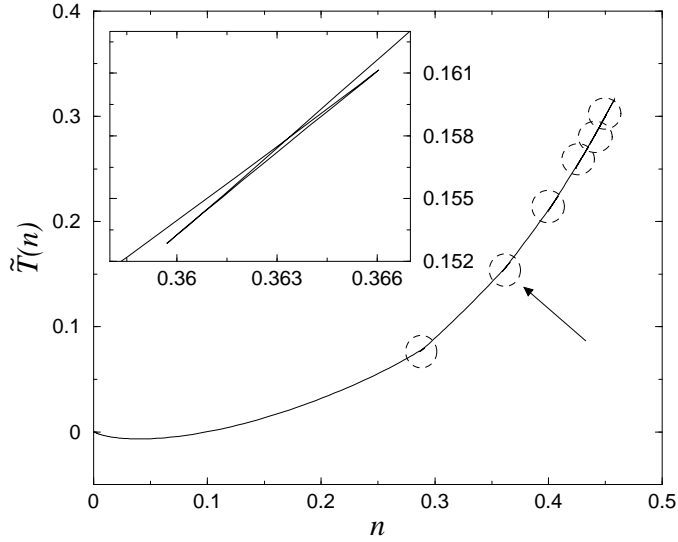,width=0.5\linewidth}
\caption{Local kinetic energy density (in units of 
$N^{-3/2}\hbar \omega/a_{ho}$)
as a  function of the particle
density (in units of $ N^{-1/2} a_{ho}^{-1}$) for 
$N=12$.
The circles indicate the regions in which the function is
multi-valued. 
The inset shows an enlargement of one of these regions, indicated by
the arrow. The
function $\tilde T(n)$ is multi-valued around all 
local maxima 
of the density profile.}
\label{fig4}
\end{figure}

Knowing exactly the function $T(x)$  the specific case of
 1D harmonic confinement, we may try to construct a function
$\tilde T (n)=T(x(n))$  in a local density approach, with
 $x(n)$ being given by inversion  of the density profile. The
inversion has to be performed locally, since the exact
 density profile 
is not globally invertible owing to its shell structure. The
resulting $\tilde T(n)$, as reported  in Fig~\ref{fig4}, is a
multi-valued function around the values of the particle density corresponding
to each  local maximum in the  profile.
While this shows that a local-density approach cannot 
characterize the operator $\hat T$, further progress can be made by
studying the derivative of $\tilde T(n)$ with respect to the density,
 which is the local 
chemical potential $\tilde \mu_{loc}(n)=d\tilde T(n)/dn$.

We show here that for the specific case of a 1D harmonic confinement the
inverse $\tilde n (\mu_{loc})$ of the function $\tilde \mu_{loc}(n)$ is a
single-valued function. This is most simply seen by inverting the Euler
equation 
\begin{equation}
\frac{\delta E_{kin}[n(x)]}{\delta n(x)}=\mu-V_{ext}(x)\equiv
\mu_{loc}(x)
\label{muloc}
\end{equation}
in the domain $x\ge 0$ 
and by substituting its value into the density profile $n(x)$ 
to obtain
\begin{equation}
\tilde n(\mu_{loc})=n(x(\mu_{loc}))\;.
\end{equation}
Here, $x(\mu_{loc})$  is the inverse of the function
$\mu_{loc}(x)$. Because of the symmetry $x \leftrightarrow -x$ for the
harmonic potential, the region $x<0$ is not needed to obtain
$\tilde n(\mu_{loc})$. 

\begin{figure}
\epsfig{file=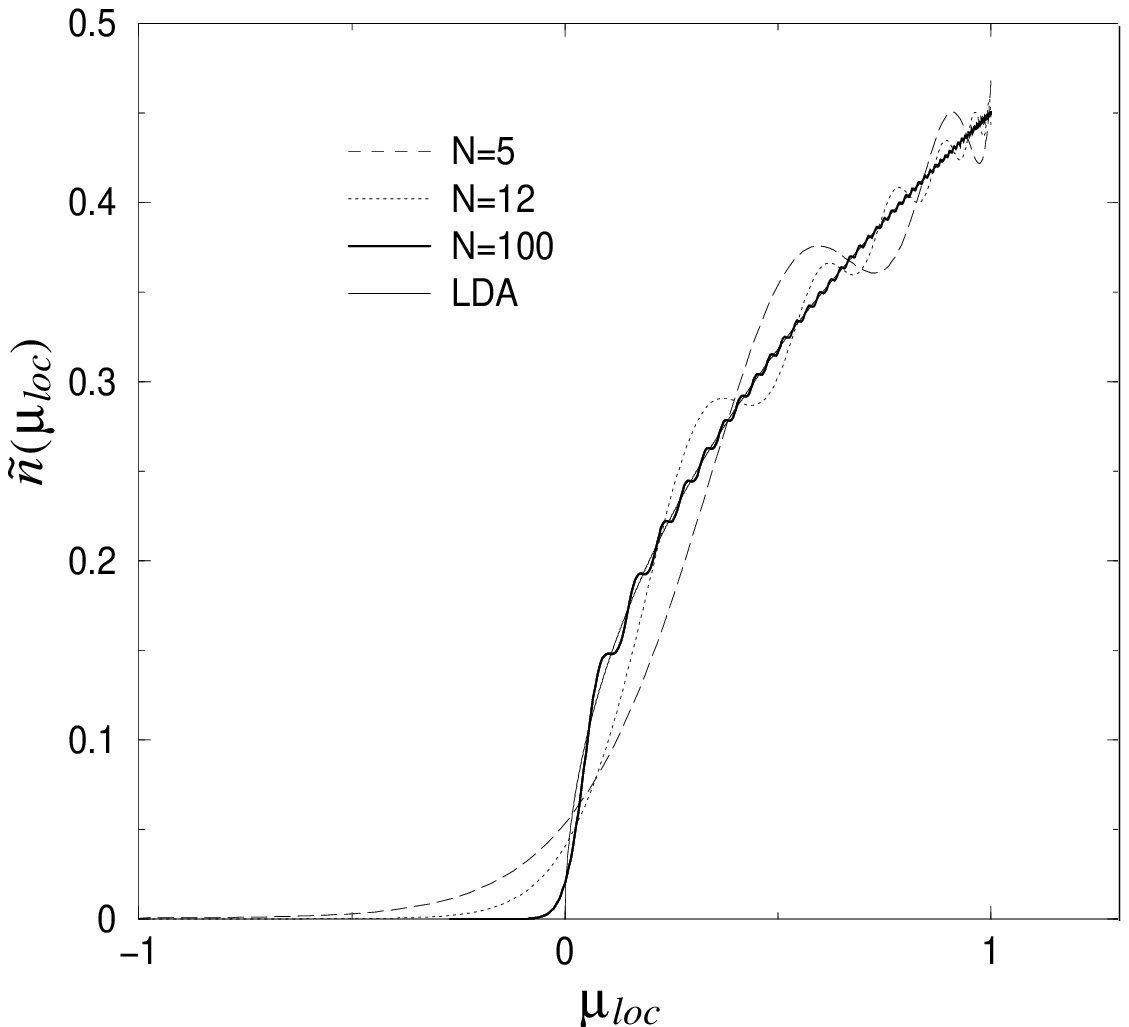,width=0.5\linewidth}
\caption{Particle density $\tilde n$ (in units of $N^{-1/2} a_{ho}^{-1}$)
 as a  function of ${\mu}_{loc}$ (in units of $\hbar\omega/N$)
for various  values of $N$ as compared with the LDA profile.}
\label{fig5}
\end{figure}

An analytic calculation  for $\tilde n(\mu_{loc})$ has been reported by
Lawes and March \cite{march} for $N=1$ and 2. The numerical method described in
Sect.~\ref{sec1} allows us to easily  compute the density profile for
numbers of particles up to $N=1000$. By combining these results for
$n(x)$ with the analytic expression for $x(\mu_{loc})$, we have
obtained the function $\tilde n(\mu_{loc})$ which is shown in
Fig.~\ref{fig5} for various values of $N$. At variance from the LDA
prediction, the exact density profile is non-vanishing at negative
values of $\mu_{loc}$ and shows a number of oscillations (equal 
 to  $N/2$ for even $N$ or to $(N-1)/2$ for odd $N$), with 
 decreasing amplitude as $N$  increases. 

The property of single-valuedness of $\tilde n(\mu_{loc})$ is a
 characteristic of
1D harmonic confinement and can be understood to be  a consequence of the
local nature of the relation $\tilde n(\mu_{loc})$ {\it
 vs.}~$\mu_{loc}$, which was 
already pointed out by Lawes and March  \cite{march}. Anharmonicity
 would be sufficient to make it invalid.

\section{Summary and conclusions}
\label{secconcl}
In summary, we have presented a general method for evaluating
the $n$-th moments of the one-body density matrix (defined either
through the derivatives with respect to the relative coordinate 
or through the derivatives 
 with
respect to the second position variable)
for a system of confined non-interacting 1D fermions.
We have applied this method to the case of  harmonic 
confinement to calculate up to fourth moments.

Regarding  second moments, the momentum flux density $\Pi(x)$
has been
computed and compared with the 
kinetic energy density $T(x)$ at various numbers of particles. While
$T(x)$ shows a prominent shell structure and negative
tails, $\Pi(x)$ is everywhere positive and has a
less marked 
shell structure, which can be  evidenced in its first spatial 
derivative. The relation between these two different physical
quantities has been 
elucidated by giving an exact relation between them  and
the equilibrium density profile of the fermion cloud. The momentum flux
density determines
 the dynamical properties of
the cloud, whereas the kinetic energy density is relevant in the
context of 
Density Functional Theory. 

We have also calculated the fourth moments of the  density
matrix in order to display the quantal fluctuations of the kinetic
energy and  momentum flux densities. Finally, we have evaluated the
local relationship which exists between density profile and local
chemical potential for  fermion clouds in 1D harmonic confinement.
This relation has allowed us to display an exact property of the DFT
kinetic energy functional for a 1D Fermi gas up to mesoscopic values
of the number of particles.

\acknowledgements
This work was partially supported by MURST through PRIN2000.
One of us (A.M.)
would like to thank Dr. F. Illuminati for useful discussions and
for giving reference to the work of Ziff {\it et al.}  \cite{uhlenbeck}.

\appendix
\section{Density profile from the Kirkman-Pendry relation}
We derive in this Appendix an expression alternative to
Eq.~(\ref{penta})  for the evaluation of the density profile in a 
 system of $N$ fermions under 1D 
 harmonic confinement. 
The trace of the Green's function $\hat
G(x)$ over its first N states in Eq.~(\ref{equno})
is obtained through an extension 
 of the
Kirkman-Pendry relation  \cite{pendry},
 already used for  the density of
states in a  quasi-1D system
 \cite{vfg,fgv}.

We define the Green's function
\begin{equation}
\label{mancava}
{\cal \hat G}(\delta,x)=\dfrac{1}{x+\delta-\hat \xi(x)+i\varepsilon}
\end{equation}
where $\delta$ is an auxiliary continuous
 variable
and $\hat \xi(x)$ is an  effective position operator of dimension $N\times
N$. This  is defined by setting  $[\hat\xi(x)]_{i,j}=[\hat x]_{i,j}$ if
$(i,j)\neq(N,N)$ and $[\hat\xi(x)]_{N,N}=\tilde x_{N,N}(x)$, with
\begin{equation}
\tilde{x}_{N,N}(x)=\cfrac{N/2}{x+i\varepsilon-\cfrac{(N+1)/2}
{x+i\varepsilon-\dots}}\;.
\label{contfr}
\end{equation}
The
single term $\tilde x_{N,N}(x)$ contains 
the contribution 
of all the states  which are not occupied by the fermions.
Its asymptotic value is $\tilde x_{N,N}(x)=i\sqrt{2N-x^2}$  for $N\rightarrow \infty$.

An expression for ${\rm Tr}_N\,{\hat G(x)}$ is obtained from
the Green's function element ${\cal G}_{1,N}(\delta,x)$
between the first and the last occupied
state by using the expression
\begin{eqnarray}
\dfrac{\partial}{\partial \delta}\ln {\cal G}_{1,N}(\delta,x)
&=&
\dfrac{\partial}{\partial \delta}\ln
\dfrac{\prod_{i=1}^{N-1}[\hat \xi(x)]_{i,i+1}}
{\det[x+\delta-\hat\xi(x)+i\varepsilon]}
=
\dfrac{\partial}{\partial \delta}\ln
\dfrac{\prod_{i=1}^{N-1}[\hat\xi(x)]_{i,i+1}}
{\prod_{i=1}^{N}{[x+\delta-\xi_i(x)+i\varepsilon]}}\nonumber\\
&=&
\dfrac{\partial}{\partial \delta}\ln \prod_{i=1}^{N} 
\dfrac{1}{x+\delta-\xi_i(x)+i\varepsilon}
=
\sum_{i=1}^{N}
\dfrac{-1}{x+\delta-\xi_i(x)+i\varepsilon}
\end{eqnarray}
where $\xi_i(x)$ are the eigenvalues of the operator
$\hat\xi(x)$.
Therefore, the particle density is given by
\begin{equation}
n(x)=-\dfrac{1}{\pi}\lim_{\varepsilon\rightarrow 0^+}{\rm Im}{\rm Tr}_N
\hat G(x)= 
\dfrac{1}{\pi}\lim_{\varepsilon\rightarrow 0^+} {\rm Im}\,\left[\dfrac{\partial}{\partial \delta}
\ln {\cal G}_{1,N}(\delta,x)
\right]_{\delta=0}\,.
\label{SPERO}
\end{equation}
The matrix element ${\cal G}_{1,N}(\delta,x)$ can be calculated by a further
renormalization \cite{vfg} of $\hat\xi(x)$, without explicitly 
inverting
 the matrix $(x+\delta-\hat\xi(x)+i\varepsilon)$.
As in 
 the method presented in  Sec.~\ref{sec1},
convergence is achieved by setting
$M=10^7$ and $\varepsilon=0.001$.

The same procedure can be used to evaluate the momentum distribution
as the trace of an  $N\times N$  operator 
$\hat\pi(p)$, defined analogously to $\hat \xi(x)$.
Due to the symmetry between
the operators $\hat{x}$ and $\hat{p}$ in the harmonic oscillator
Hamiltonian, we obtain essentially 
the same final  expression as for
of the particle density $n(x)$.
Thus, the momentum distribution for $N$ fermions under 1D harmonic
confinement shows 
$N$ oscillations.

The 
mapping  between hard-core bosons and
non-interacting fermions in 1D  \cite{girardeau} 
allows us to evaluate the particle density profile for mesoscopic
systems of strongly correlated bosons by the same method. However, 
the
mapping is 
restricted to real-space properties and 
does not apply to the momentum distribution.

\begin{thebibliography}{10}
\bibitem{fermioni} B. DeMarco and D.\,S. Jin, Science {\bf 285}, 1703 (1999).

\bibitem{bec_rb}
M.~H. Anderson, J.~R. Ensher, M.~R. Matthews, C.~E. Wieman, and  E.~A. Cornell,
  Science {\bf 269},  198  (1995).

\bibitem{bec_na}
K.~B. Davis, M.~O. Mewes, M.~R. Andrews, N.~J. van Druten, D.~S. Durfee, D.~M.
  Kurn, and  W. Ketterle, Phys. Rev. Lett. {\bf 75},  3969  (1995).

\bibitem{bec_li}
C.~C. Bradley, C.~A. Sackett, J.~J. Tollet, and R.~G. Hulet,
Phys. Rev. Lett. {\bf 75}, 1687 (1995); see also {\it ibid.}
{\bf 79}, 1170 (1997).

\bibitem{lund} S. Lundqvist and N.~H. March, {\em Theory of the Inhomogeneous
Electron Gas} (Plenum, New York, 1983); E.~K.~U. Gross, J.~K. Dobson,
and M. Petersilka, in {\em Topics in Current Chemistry},
ed. R.~F. Nalewajski (Springer, Berlin, 1996), p.~1. 

\bibitem{noi}
P. Vignolo, A. Minguzzi, and M.~P. Tosi, Phys. Rev. Lett. {\bf 85},
2850 (2000). A misprint has occurred  in the first line of
Eq.~(3), which should read $P(x)=-(\hbar^2/m) (\partial^2/\partial x_1^2)
\rho(x,x_1)|_{x_1=x}$. However, the second line of Eq.~(3) for $P(x)$ 
 and all the following calculations are correct.

\bibitem{schneider}
J. Schneider and H. Wallis, Phys. Rev. A {\bf 57}, 1253 (1998). 

\bibitem{bruun}
G.~M. Bruun and K. Burnett, Phys. Rev. A {\bf 58}, 2427 (1998). 

\bibitem{zimmermann}
F. Gleisberg, W. Wonneberger, U. Schl\"oder and C. Zimmermann,
Phys. Rev. A {\bf 62}, 063602 (2000). 

\bibitem{girardeau}
M.~D. Girardeau, Journ. Math. Phys. {\bf 1}, 516 (1960).

\bibitem{girardnew}
M.~D. Girardeau and E.~M. Wright, Phys. Rev. Lett. {\bf 84}, 5239 (2000).

\bibitem{kolomeisky}
E.~B. Kolomeisky, T.~J. Newman, J.~P. Straley, and X. Qi, Phys. Rev. Lett.
{\bf 85}, 1146 (2000).

\bibitem{shlyap}
D.~S. Petrov, G.~V. Shlyapnikov, and J.~T.~M. Walraven, cond-mat/0006339.  

\bibitem{uhlenbeck}
R.~M.~Ziff, G.~E.~Uhlenbeck, and M.~Kag, Phys. Rep. {\bf 32}, 
169 (1977).

\bibitem{vfg}
P. Vignolo, R. Farchioni, and G. Grosso, Phys. Rev. B {\bf59},  16065 (1999).

\bibitem{fvg}
R. Farchioni, P. Vignolo, and G. Grosso, Phys. Rev. B {\bf60}, 15705 (1999).

\bibitem{fgv}
R. Farchioni, G. Grosso and P. Vignolo, Phys. Rev. B {\bf 62}, 12565
(2000).


\bibitem{march}
G.~P. Lawes and N.~H. March, J. Chem. Phys. {\bf 71}, 1007 (1979).

\bibitem{noticina} This condition is not restrictive since in 1D we can
use the non-degeneration theorem [L.~D. Landau and E.~M. Lifshitz, {\em Quantum Mechanics: Non
Relativistic Theory} (Pergamon, Oxford 1959), p.57] to show that all 
eigenfunctions 
vanishing at infinity are real.

\bibitem{pendry}
P.~D. Kirkman and J.~B. Pendry, J. Phys. C {\bf 17}, 4327 (1984).

\bibitem{march_tosi}
N.~H. March and M.~P. Tosi, Proc. R. Soc. Lond. A {\bf 330}, 373 (1972).


\bibitem{rinor}
G. Grosso and G. Pastori Parravicini, {\it Solid State Physics} 
(Academic, London 2000) p.~191.

\end{thebibliography}
\end{document}